\documentclass[12pt,a4paper]{article}

\usepackage[latin1]{inputenc}
\usepackage[english]{babel}

  \usepackage{a4wide}
  \usepackage{amsmath,amssymb,amsthm}
\renewcommand{\d}{\textrm{d}}
\newcommand{\Real}{\textrm{I\!R}}
\newcommand{\e}{\textrm{e}}
\newcommand{\SU}{\mathop{\rm SU}}
\newcommand{\SO}{\mathop{\rm SO}}
\newcommand{\SL}{\mathop{\rm SL}}

\newcommand{\ve}{\epsilon_*}

\begin{document}

\begin{flushright}
\small UUITP-01/10\\

\date \\
\normalsize
\end{flushright}

\begin{center}

\vspace{.5cm}

{\LARGE \bf{Lifshitz backgrounds from 10d supergravity}} \\
\vspace{0.2cm}

\vspace{1.5 cm} {\large  Johan Bl{\aa}b\"ack, Ulf H.~Danielsson and Thomas Van Riet }\\
\vspace{0.6 cm}  {Institutionen f\"or fysik och
astronomi\\
Uppsala Universitet, Box 803, SE-751 08 Uppsala, Sweden}\\
\vspace{0.4cm} {\upshape\ttfamily johan.blaback, ulf.danielsson,
thomas.vanriet@fysast.uu.se} \\

\vspace{1.5cm}

{\bf Abstract}
\end{center}

\begin{quotation}
\small We investigate whether 4-dimensional static and cosmological
Lifshitz solutions can be found from deforming the existing
(A)dS$_4$ compactifications in IIA and IIA$^{\star}$ supergravity.
Using a well motivated compactification Ansatz on $\SU(3)$-structure
manifolds with 19 undetermined parameters we demonstrate that this
is not the case in ordinary IIA supergravity, thereby generalising
previous nogo results in different ways. On the other hand, for
IIA$^*$ we construct explicit cosmological Lifshitz solutions. We
also consider solutions with non-constant scalars and are able to
find simple static and cosmological Lifshitz solutions in IIB$^{*}$
supergravity and a Euclidean Lifshitz solution in ordinary Euclidean
IIB supergravity, which is similar to a non-extremal deformation of
the D-instanton. The latter solutions have $z=-2$.
\end{quotation}

\newpage

\pagestyle{plain}

%\tableofcontents

\section{Introduction}
Since the advent of the gravity/Lifshitz-QFT correspondence
\cite{Kachru:2008yh} there is the need for an embedding of Lifshitz
spacetime ($Li$) in string theory. Such an embedding might allow for
a microscopic understanding of the correspondence and allows one to
define the correspondence for all values of the gravitational
coupling since string theory is a UV complete gravity theory. There
is also a more practical use if the embedding can be done in the
supergravity limit. This would allow one to investigate possible
supersymmetry preservation and hence stability issues.

Recently, reference \cite{Hartnoll:2009ns} found some stringy
constructions of Lifshitz spacetime but they seem to fail to be
simple supergravity solutions, and it is not clear to us whether the
solutions are really 10-dimensional solutions. A perhaps simpler
approach would be to consider supergravity flux-compactifications to
4 dimensions for which the effective theories have the ingredients
required for supporting Lifshitz spacetimes: massive vectors or
massive tensors and a negative cosmological constant. The existence
of these ingredients is not enough to guarantee a solution. This
usually comes from the complicated scalar field interactions in the
effective theory \cite{Taylor:2008tg}. Insisting on fixed scalar
fields, when vectors are turned on, is a non-trivial problem and
resembles the issue of the attractor mechanism of black hole
solutions in string theory. In this paper we will take the scalar
fields constant and we comment in the end that this might be the
reason behind the various nogo conditions we find.

Reference \cite{Azeyanagi:2009pr} was able to construct a Lifshitz
solution from a flux compactification of IIB supergravity (based on earlier work \cite{Fujita:2009kw}). This
solution has a non-constant dilaton field, but a more pressing
problem is the fact that the Lifshitz anisotropy is only possible in
one direction. This is an unwanted feature in the correspondence as
was later argued by the same authors in \cite{Li:2009pf}. For that
reason an investigation was performed for flux compactifications of
11- and 10-dimensional supergravity under specific Ansatze for which
nogo theorems could be proven. In the present paper we continue this
investigation but specify to the case of (massive) IIA
compactifications on $\SU(3)$-structure manifolds with fluxes and
with the possibility of wrapped O6/D6 sources. These
compactifications are known for its susy AdS$_4$ vacua with the
possibility of stabilising all moduli at tree-level in the string
coupling $g_s$ and $\alpha'$, see \cite{DeWolfe:2005uu, Lust:2004ig,
Caviezel:2008ik} and references therein. The effective theories are
expected to possess massive vectors, aside from the already present
possibility of minimizing the potential at a negative value.

The $\SU(3)$-structure manifolds that support the susy AdS$_4$
solutions have 2 of the 5 torsion classes non-zero. In this setup we
make the most general Ansatz for the fluxes that can be studied in
all generality without specifying to a specific model. This implies
that the Ansatz consists out of the so-named `canonical' forms,
being the (would-be) K\"ahler form $J$, the complex holomorphic
three-form $\Omega$ and the torsion forms $W_1, W_2$. A rationale
for such an Ansatz is the fact the internal part of the Einstein
equations seems to indicate that solutions are naturally carried by
these forms (though not necessarily). This is because the geometry
side of the Einstein equation is written entirely in terms of the
canonical forms and the same must be true for the matter side.
Indeed, the known susy AdS solutions are captured by it, as well as
the recently discovered non-susy AdS solutions in
\cite{Danielsson:2009ff} and \cite{Koerber:2010rn} (see also \cite{Cassani:2009ck} and \cite{Lust:2008zd} for related work). In
\cite{Danielsson:2009ff}, the same Ansatz was also used to establish
de Sitter solutions\footnote{Although manifolds with the desired
properties are still to be constructed. }.

An investigation of this kind was carried out in \cite{Li:2009pf}
but with $W_2=0$ and a negative result was found. We will
demonstrate that the Ansatz is much more general when $W_2$ is added
and includes \emph{19 undetermined parameters}. One would then think
that it is much easier to find Lifshitz solutions, but, as we will
demonstrate, a negative result is found again.

Just as the AdS/CFT correspondence can be generalised to the (more
hypothetical) dS/ECFT correspondence \cite{Strominger:2001pn}, the
Li/condensed matter (Li/CM) correspondence can be generalised to a
\emph{time-dependent} Li/ECM correspondence \cite{Nakayama:2009gi},
where `E' stands for Euclidean. These time-dependent Lifshitz
spacetimes will be named \emph{cosmological} Lifshitz spacetimes in
this paper. It was shown in \cite{Nakayama:2009gi} that these
spacetimes violate the null energy condition, and therefore the
supporting matter content should be non-trivial. One possibility
that was mentioned in \cite{Nakayama:2009gi} is the use of
orientifolds as null-energy breaking ingredients. In this paper we
allow for orientifolds in IIA supergravity and search for
cosmological Lifshitz solutions without success. Nonetheless, these
compactifications we consider are known to allow de Sitter critical
points \cite{Flauger:2008ad,
Caviezel:2008tf, Danielsson:2009ff}.% This is probably a necessary
%condition for having cosmological Lifshitz solutions, just as AdS
%solutions probably are a necessary condition in Lifshitz
%compactifications.

Reference \cite{Nakayama:2009gi} also suggested ghostlike matter as
a possibility. Such matter can be introduced into 10d supergravity
in two ways: 1) Euclidean IIB is known to have the flipped sign for
the RR zero-form kinetic term. 2) The so-called II$^*$ theories are
obtained from timelike T-duality of the ordinary type II theories
and have the flipped sign for all RR kinetic terms
\cite{Hull:1998fh, Hull:1998vg}. On of the properties of the star
theories is that the natural vacua are de Sitter instead of Anti-de
Sitter. Timelike T-duality is a contested concept for various
reasons and one of them is exactly the ghostlike matter fields. But
as explained in \cite{Nakayama:2009gi} this might be turned into a
virtue since Euclidean CM theories can break unitarity. We
demonstrate explicitly that the ghostlike terms in IIA$^*$
supergravity indeed allow cosmological Lifshitz solutions!

Finally we initiate in this paper the search for Lifshitz solutions
carried by non-constant scalar fields that depend on the holographic
coordinate. We demonstrate the existence of Euclidean Lifshitz
solutions in Euclidean IIB and static and cosmological Lifshitz
solutions in IIB$^*$. The reason for this is exactly the flipped
sign in the kinetic terms for the RR zero form. Note that, whereas
IIB$^*$ might be contested, this is not the case for Euclidean IIB
theory. It is known that the usual AdS/CFT correspondence can be
described either in the Euclidean or the Lorentzian way
\cite{Balasubramanian:1998de}.

The rest of this paper is organised as follows. In section 2 we
describe some basic features of the Lifshitz geometries required for
the rest of the paper. In section 3 we investigate the deformations
of the (A)dS$_4$ flux compactifications on $\SU(3)$-structure spaces
of IIA and IIA$^*$ theory. In section 4 we consider Euclidean IIB
and IIB$^*$ theory with running axion-dilaton scalars and present
the Lifshitz solutions. Finally in section 5 we discuss the obtained
results and further directions for research.

\section{Lifshitz spaces: static, cosmological and Euclidean}
Consider the following three line-elements of 4-dimensional Lifshitz
spaces (Li$_4$)
\begin{align}
&\d s_s^2= -\frac{\d x_0^2}{r^{2a}} + \frac{\d x_1^2}{r^{2b}}  + \frac{\d x_2^2}{r^{2c}} +\frac{\d r^2}{r^2} \,,\\
&\d s_c^2=+\frac{\d x_0^2}{r^{2a}} + \frac{\d x_1^2}{r^{2b}}  +
\frac{\d x_2^2}{r^{2c}} -\frac{\d r^2}{r^2} \,,\\
&\d s_e^2= +\frac{\d x_0^2}{r^{2a}} + \frac{\d x_1^2}{r^{2b}}  +
\frac{\d x_2^2}{r^{2c}} +\frac{\d r^2}{r^2} \,,
\end{align}
where the subscript `s' means static, `c' means cosmological and `e'
means Euclidean. In the static case $x_0$ is the time direction and
in the cosmological case $r$ is the time direction. For each example
the holographic coordinate is denoted by $r$ (the RG scale) and the
dual field theory lives on the slice of constant $r$. When we
rescale $r\rightarrow \lambda r$ then the metric is invariant if
\begin{equation}
x_0\rightarrow \lambda^{a}x_0\,,\qquad x_1\rightarrow
\lambda^{b}x_1\,,\qquad x_2\rightarrow \lambda^{c}x_2\,.
\end{equation}

In many cases it is desired to treat two directions on the
holographic slice as ``equal'' by taking $b=c$. We can then find a
coordinate transformation such that $b=c=-1$. If we furthermore
denote $a=-z$, the line elements become
\begin{align}
&\d s_s^2= -r^{2z}\d x_0^2 + r^{2}(\d x_1^2  + \d x_2^2) + \frac{\d r^2}{r^2} \,,\\
&\d s_c^2= +r^{2z}\d x_0^2 + r^{2}(\d x_1^2  + \d x_2^2) - \frac{\d r^2}{r^2} \,,\\
&\d s_e^2= +r^{2z}\d x_0^2 + r^{2}(\d x_1^2  + \d x_2^2) + \frac{\d
r^2}{r^2}\,.
\end{align}
$z$ is  the ``anisotropy'' parameter. When $z=1$ we have (A)dS$_4$
space. For this reason we call $z\neq 1$ anisotropic, since time and
space scale differently. For original work on Lifshitz spacetimes we refer to \cite{Koroteev:2007yp}.

The Vielbein one-forms are denoted $\theta^0=r^z\d x_0,
\theta^r=r^{-1}\d r,\theta^{1,2}=r\d x^{1,2}$ in accordance with the
existing literature. They obey the following simple Cartan--Maurer
equations
\begin{equation}
\d\theta^r=0\,,\qquad\d \theta^0=z\theta^r\wedge\theta^0\,,\qquad
\d\theta^{1,2}=\theta^r\wedge\theta^{1,2}\,.
\end{equation}
For later use we present the Ricci curvatures in the general case
\begin{align}
&R^s_{00}=-R^e_{00}=+R^c_{00}=+a(a+b+c)r^{-2a}\,,\nonumber\\
&R^s_{11}=+R^e_{11}=-R^c_{11}=-b(a+b+c)r^{-2b}\,,\nonumber\\
&R^s_{22}=+R^e_{22}=-R^c_{22}=-c(a+b+c)r^{-2c}\,,\nonumber\\
&R^s_{rr}=+R^e_{rr}=+R^c_{rr}=-(a^2+b^2+c^2)r^{-2}\,.
\end{align}

The generalisation of all the above to any dimension is
straightforward.

\section{A nogo for some massive IIA compactifications}

\subsection{Type IIA/IIA$^*$ supergravity}
%The action for type II supergravity theories, up to Chern--Simons
%terms, are (where we have put $\kappa^2_{10}$=1/2)
%\begin{equation}
%S_{bulk}=\int\sqrt{g}\Bigl\{R-\tfrac{1}{2}(\partial\phi)^2-\sum_n\frac{1}{2\,n!}\e^{a_n\phi}F_n^2\Bigr\}\,,
%\end{equation}
%where $\Sigma_n$ represents the sum over all the field strengths and
%the numbers $a_n$ are given by
%\begin{equation}
%a^{RR}_n=\frac{5-n}{2}\,,\qquad a^{NS}_3=-1\,.
%\end{equation}
In IIA and IIA$^{*}$ theory the RR field strengths are $F_0, F_2,
F_4$. The difference in the IIA$^{*}$ theory is that the RR field
strengths are transformed according to $F_p\to iF_p$.  We implement
this by adding a sign, $\ve$, such that $\ve=1$ corresponds to type
IIA and $\ve=-1$ is type IIA$^{*}$. The form and dilaton equations
of motion in Einstein frame are
\begin{align}
& \d (\star\e^{3\phi/2} F_{2}) + \e^{\phi/2}\star F_{4}\wedge H=0\,,\nonumber \\
& \d (\star \e^{\phi/2} F_{4})- F_{4}\wedge H=0\,,\nonumber \\
& \ve\d(\star\e^{-\phi}H)  + \e^{\phi/2}\star F_4\wedge F_2
-\tfrac{1}{2}F_4\wedge F_4 + F_0\e^{3 \phi/2}\star F_2=0\,,\label{H eq}\\
& \ve\d\star\d\phi -\tfrac{1}{4}\e^{\phi/2}\star F_4\wedge F_4 +
\ve\tfrac{1}{2}\e^{-\phi}\star H\wedge H -
\tfrac{3}{4}\e^{3\phi/2}\star F_2\wedge F_2 -
\tfrac{5}{4}\e^{5\phi/2}\star F_0\wedge F_0=0\,,\nonumber
\end{align}
where $F_0$ is the Romans' mass $m$. The Bianchi identities read
\begin{equation}\label{Bianchi}
\d H_3=0\,,\qquad \d F_2=F_0H\,,\qquad\d F_4=F_2\wedge H_3\,.
\end{equation}
The Einstein equation is given by
\begin{align}
&0=\ve\mathcal{R}_{ab}-\tfrac{1}{2}\ve\partial_{a}\phi\partial_{b}\phi-\tfrac{1}{12}\e^{\phi/2}F_{acde}F_b^{\,\,cde}
+ \tfrac{1}{128}\e^{\phi/2}g_{ab}F_4^2 -
\ve\tfrac{1}{4}\e^{-\phi}H_{acd}H_{b}^{\,\,cd} \nonumber\\& +
\ve\tfrac{1}{48}\e^{-\phi}g_{ab}H^2
-\tfrac{1}{2}\e^{3\phi/2}F_{ac}F_b^{\,\,c}
+\tfrac{1}{32}\e^{3\phi/2}g_{ab}F_2^2-
\tfrac{1}{16}g_{ab}\e^{5\phi/2}F_0^2\,.\label{Einstein eq}
\end{align}

\subsection{Spaces with $SU(3)$-structure}
A $\SU(3)$-structure space is characterised by a real two form $J$
and a complex three form $\Omega=\Omega_R + i\Omega_I$. The exterior
derivatives are given by
\begin{align}
&\d J = -\frac{3i}{2}W_1\Omega_R + W_3 + W_4\wedge J\,,\nonumber\\
&\d\Omega = W_1 J\wedge J + W_2\wedge J+ W_5\wedge \Omega\,,
\end{align}
where the $W_i$ are complex forms whose rank can be deduced from the
above equations. In the following we restrict to spaces for which
$W_3=W_4=W_5=0$ and where $W_1$ is an imaginary zero-form and $W_2$
is an imaginary two-form for reasons explained in the introduction.
It is expected that, as the moduli flow away from the susy AdS
solution, other torsion classes can be turned on. Nonetheless we
will consider the case were only these two torsion classes are
non-zero.

These forms obey the following form identities
\begin{align}
& \star_6\Omega=-i\Omega\,,\qquad\star_6 J=\tfrac{1}{2}\,J\wedge J \,,\qquad \star_6W_2=-J\wedge W_2\,,\nonumber\\
& \Omega\wedge\Omega^*=\tfrac{4i}{3}\,J\wedge J\wedge J\,,\qquad
J\wedge J\wedge J= 6 \epsilon_6\,,\nonumber\\
& \Omega\wedge J=0\,,\qquad W_2\wedge J\wedge J=0\,,\qquad W_2\wedge
\Omega=0\,,
\end{align}
and the following contractions\footnote{ The notation we use for
``squaring'' a tensor $T_{i_1\ldots i_n}$ is $
T^2_{ij}=T_{ii_2i_3\ldots i_n}T_j^{\,\,i_2i_3\ldots i_n}\,. $}
\begin{align}
& J_{mn}W_2^{mn}=0\,,\qquad
J_{m}^{\,\,\,n}J_p^{\,\,\,q}(W_2)_{nq}=(W_2)_{mp}\,,\nonumber\\
& (\Omega_R)^2_{ab}=(\Omega_I)^2_{ab}=4g_{ab}\,,\qquad
J^2_{ab}=g_{ab}\,.
\end{align}
We furthermore assume that $\d W_2$ is proportional to $\Omega_R$.
The constant of proportionality is fixed by internal consistency
\begin{equation}
\label{dW2} \d W_2=-(i|W_2|^2/8)\, \Omega_R\,.
\end{equation}
This condition is rather common for many explicit geometries
\cite{Caviezel:2008ik} and is required for some Ad$S_4$
\cite{Lust:2004ig} and  dS$_4$ solutions \cite{Danielsson:2009ff}.

For the sake of solving the equations of motion, we single out a
special class of geometries which we call `degenerate' since some
tensors become linearly dependent on each other
\begin{equation}
W_2 \wedge W_2 = \frac{1}{12}|W_2|^2J\wedge J  -2i\chi J\wedge
W_2\quad\Longrightarrow\quad (W^2_2)_{ij} = \frac{W_2^2}{6}g_{ij} +
i\chi (JW_2)_{ij}\,,\label{degenerate}
\end{equation}
with $\chi$ some real number different from zero. The degenerate
condition is a necessary condition for having  non-susy (A)dS
solutions in this setup as described in \cite{Danielsson:2009ff,
Koerber:2010rn}.

One can also express the curvature tensors in terms of the torsion
classes \cite{Ali:2006gd, bedulli-2007-4}
\begin{equation}
\mathcal{R}_{mn}=-\frac{3i}{4}(\Omega_R)_n^{\,\,ps}\partial_{[p}(W_2)_{sm]}
-\frac{1}{4}W_1(W_2)_{mr}J_{n}^{\,\,\,r} -
\frac{1}{2}(W_2)_{mq}(W_2)_{n}^{\,\,\,q}
+\frac{5}{4}g_{mn}|W_1|^2\,.
\end{equation}

\subsection{The Ansatz}
We consider Lifshitz spacetimes with just one anisotropy parameter
($z$) and we add the symbol $\epsilon$ to distinguish between the
static ($\epsilon=+1$) and cosmological ($\epsilon=-1$) case
\begin{equation}
\d s^2= -\epsilon r^{2z}\d x_0^2 + r^{2}(\d x_1^2  + \d x_2^2) +
\epsilon \frac{\d r^2}{r^2} \,.
\end{equation}
The Euclidean case will be considered later. This $\epsilon$ is not
be confused with the $\ve$ that distinguishes between normal and
star supergravity.

At this point we can make the most general Ansatz consistent with
the 4-dimensional Lifshitz symmetries and which features fluxes
along the canonical forms $J, W_2,\Omega$ and wedges thereof
\begin{align}
& F_2=a J + \alpha\theta^0\wedge\theta^r+\eta\theta^1\wedge\theta^2 + icW_2 \,,\nonumber\\
& H_3=\beta\theta^1\wedge\theta^2\wedge\theta^r + k\Omega_R\,,\nonumber\\
&F_4=f\theta^0\wedge\theta^1\wedge\theta^2\wedge\theta^r +
g\theta^1\wedge\theta^2\wedge J +h \theta^0\wedge\theta^r\wedge J +
q\theta^0\wedge\Omega_R \nonumber\\
&\qquad +\frac{s}{2}J\wedge J +ie\theta^1\wedge\theta^2\wedge W_2 +
il\theta^0\wedge\theta^r\wedge W_2 + ipW_2\wedge J\,.
\end{align}
We have eliminated terms from the most general Ansatz with canonical
forms which obviously have to be zero from Bianchi identities and
form equations of motion, such as $\theta^0\wedge \Omega_I$.

When we take the Lorentz symmetry-breaking terms equal to zero
\begin{equation}
\alpha=\eta=\beta=g=h=q=e=l=0\,,
\end{equation}
we end up with the Ansatz used for (non)-susy (A)dS solutions.
Hence, this is the natural Ansatz that is expected to ``deform'' the
(A)dS solutions into Lifshitz solutions. As an example, the susy AdS
solutions are given by
\begin{align}
&a=\tfrac{1}{4}iW_1\,,\qquad k=-\tfrac{2}{5}m\,,\qquad
f=\tfrac{9}{4}iW_1\,,\qquad s=\tfrac{3}{5}m\,,\qquad c=1\,,
\end{align}
with the extra conditions that
\begin{align}
& |W_2|^2=3|W_1|^2 -\tfrac{16}{5}m^2\,,\nonumber\\
& 6=\tfrac{27}{8}|W_1|^2 +\tfrac{6}{25} m^2\,.
\end{align}
Where the second line sets the value of the cosmological constant,
which we have fixed to be $R_4=-12$.

Furthermore, we take the smeared O6/D6 sources in the usual way for
these compactifications which means \cite{Caviezel:2008ik,
Danielsson:2009ff}
\begin{align}
& \d F_2= mH + \mu\Omega_R\,,\nonumber\\
&\d\star\d\Phi=\ldots-3\mu\epsilon_{10}\,,\nonumber\\
& R_{\mu\nu}=\ldots +\frac{1}{4}\mu g_{\mu\nu}\,,\nonumber\\
& R_{ij}=\ldots -\frac{3}{4}\mu g_{ij}\,,
\end{align}
where $\mu>0$ implies net O6 charge and $\mu<0$ implies net D6
charge. Especially in the cosmological case ($\epsilon=-1$) we are
required to add orientifolds because we need to violate the
null-energy condition, as explained in \cite{Nakayama:2009gi}. In
the IIA$^{*}$ case ($\ve=-1$) there do not exist space-filling
sources since the sources of type IIA$^{*}$ have Euclidean
worldvolumes. Hence we take $\mu=0$ when $\ve=-1$.

We have all the necessary information to investigate the equations
of motion and we expect to end up with a system of many algebraic
relations in the 19 \emph{real} variables
\begin{equation}
z, iW_1, |W_2|^2, a, \alpha, \eta, c, \beta, k, f, g, h, q, s, e, l,
p, m, \mu\,.\nonumber
\end{equation}
When we plug the Ansatz into the equations of motion we need to make
a distinction between the two possible families of geometries. Let
us first assume the non-degenerate case  and later assume the
degenerate case (\ref{degenerate}).

\subsubsection*{The non-degenerate case}
The Bianchi identities give the following relations\footnote{Here we
have assumed that (\ref{dW2}) holds. If one where to relax this
condition one finds only special cases of the presented relations.}
\begin{align}
-mk-\tfrac{3}{2}iaW_1 +\tfrac{1}{8}c|W_2|^2-\mu&=0\,,\label{bianchi1}\\
-\eta k  +\tfrac{1}{8}e|W_2|^2 -\tfrac{3}{2}iW_1 g&=0\,,\label{bianchi2}\\
-\alpha k+\tfrac{1}{8}l|W_2|^2 -\tfrac{3}{2}iW_1h -q z&=0\,,\label{bianchi3}\\
m\beta-2\eta&=0\,,\label{bianchi4}\\
2g-a\beta&=0\,,\label{bianchi5}\\
2e-c\beta&=0\label{bianchi6}\,,
\end{align}
where equations (\ref{bianchi1},\ref{bianchi4}) come from the
$F_2$-field Bianchi identity and the others from the $F_4$-field
Bianchi identity. The form equations of motion give
\begin{align}
f\beta+4\epsilon qk+2\alpha&=0\,,\label{formeom1}\\
2l-\epsilon q-p\beta&=0\,,\label{formeom2}\\
\tfrac{1}{2}\beta s+h +i\epsilon qW_1&=0\,,\label{formeom3}\\
-\tfrac{3}{2}i s W_1 -\tfrac{1}{8}p|W_2|^2+\beta q -kf &=0\,,\label{formeom4}\\
el+cp&=0\,,\label{special1}\\
\epsilon \ve \beta z -\tfrac{1}{2}pl|W_2|^2 + 3hs -3ga +\alpha f
-\tfrac{1}{2}ec|W_2|^2 -m\eta&=0\,,\\
f\eta + 3ha +\frac{1}{2}lc |W_2|^2 + 3gs -\frac{1}{2}pe|W_2|^2 + m\alpha&=0\,,\label{formeom6}\\
\ve ikW_1 +\tfrac{1}{2}fs +gh -as -\tfrac{1}{2}g\eta
-\tfrac{1}{2}ma + \tfrac{1}{2}h\alpha&=0\,,\label{special2}\\
\ve k+gl+eh+fp+ap-cs-\alpha l + \eta e +mc &=0\label{special3}\,,
\end{align}
where equation (\ref{formeom1}) comes from the $F_2$ eom, equations
(\ref{formeom2}-\ref{formeom4}) come from the $F_4$ eom and all
others come from the $H$ eom. The dilaton equation of motion gives
\begin{align}
0=&\tfrac{1}{4}f^2-\tfrac{3}{4}s^2 -\tfrac{1}{8}p^2|W_2|^2-\tfrac{3}{4}g^2+\tfrac{3}{4}h^2-\tfrac{1}{8}(e^2-l^2)|W_2|^2 + \epsilon q^2 +\epsilon\ve\tfrac{1}{2}\beta^2\,,\nonumber\\
&+\ve 2k^2-\tfrac{9}{4}a^2-\tfrac{3}{8}c^2|W_2|^2+\tfrac{3}{4}\alpha^2-\tfrac{3}{4}\eta^2-\tfrac{5}{4}m^2
+3\mu\,.\label{dilaton}
\end{align}
To write the Einstein equations in a more compact form we introduce
the number $A$
\begin{align}
A=\tfrac{3}{16}f^2-\tfrac{9}{16}s^2-\tfrac{3}{32}p^2|W_2|^2-\tfrac{9}{16}g^2+\tfrac{9}{16}h^2+\tfrac{3}{32}(l^2-e^2)|W_2|^2+\epsilon\tfrac{3}{4}q^2&\,,\nonumber\\
-\epsilon\ve \tfrac{1}{8}\beta^2-\ve\tfrac{1}{2}k^2-\tfrac{3}{16}a^2-\tfrac{1}{32}c^2|W_2|^2+\tfrac{1}{16}\alpha^2-\tfrac{1}{16}\eta^2+\tfrac{1}{16}m^2&\,.
\end{align}
The external Einstein equations then read
\begin{align}
(tt)\,:\quad& \epsilon\ve z(z+2)=2\epsilon q^2 + \tfrac{1}{2}\alpha^2+\tfrac{1}{2}f^2+\tfrac{3}{2}h^2+\tfrac{l^2}{4}|W_2|^2-A-\tfrac{1}{4}\mu\,,\label{Einstein1}\\
(rr)+ (tt)\,:\quad&\ve 2(z-1)=2q^2+\ve\tfrac{1}{2} \beta^2\,,\label{Einstein2}\\
(xx)- \epsilon(rr)\,:\quad&\ve \epsilon
z(z-1)=\tfrac{1}{4}(e^2+l^2)|W_2|^2
+\tfrac{3}{2}(g^2+h^2)+\tfrac{1}{2}(\alpha^2+\eta^2)\label{Einstein3}\,.
\end{align}
Taking the trace of the 10-dimensional Einstein equation and using
the dilaton equation, we find
\begin{equation}
\epsilon
2(z^2+2z+3)=\tfrac{15}{2}|W_1|^2-\tfrac{1}{4}|W_2|^2-\tfrac{1}{2}\epsilon
\beta^2-2k^2 +2\mu\,.
\end{equation}
The traceless part of the internal Einstein then gives two more
conditions
\begin{align}
& \ve iW_1 = 4sp -4lh +4eg + 4ac  \,,\label{traceless1}\\
& \ve=-p^2 - l^2 +e^2 +c^2\,.\label{traceless2}
\end{align}
It can be shown that the Einstein equations with mixed indices are
automatically solved.  The only mixed components arise in
$(F_4^2)_{\mu i}$, and they are proportional to
$\Omega^R_{ijk}J^{jk}$ and $\Omega^R_{ijk}W_2^{jk}$. These terms
vanish since $\Omega$ is of type (3,0) and $J$ and $W_2$ are both
of type (1,1).

\subsubsection*{The degenerate case}

In the degenerate case (\ref{degenerate}) we find that conditions
(\ref{special1}), (\ref{traceless1}) and (\ref{traceless2}) do not
exist anymore and that equations (\ref{special2}) and
(\ref{special3}) are altered in the following way
\begin{align}
\ve ikW_1 +\tfrac{1}{2}fs +gh -as -\tfrac{1}{2}g\eta
-\tfrac{1}{2}ma +\tfrac{1}{2}h\alpha -\boxed{\tfrac{1}{12}(el+cp)|W_2|^2}&=0\,,\label{special2'}\\
\ve k+gl+eh+fp+ap-cs-\alpha l + \eta e +mc+\boxed{2\chi(el+cp)}
&=0\label{special3'}\,.
\end{align}
Notice that we gained one variable $\chi$ this way. However this
value gets determined by the internal Einstein equation (the
traceless part) as follows
\begin{equation}
[\ve + p^2+l^2-e^2-c^2]2\chi=iW_1\ve - 4 sp+4lh-4eg-4ac\,.
\end{equation}
This equation fixes $\chi$ and replaces the two conditions
(\ref{traceless1}) and (\ref{traceless2}).  This implies that in
total we have three equations less in the degenerate case
(\ref{degenerate}). If the degenerate case does not allow solutions
then the non-degenerate case doesn't allow solutions as well. For
that reason we will assume the degenerate case.

\subsection{Solving the equations}
\subsubsection*{No sources and nogo for cosmological $Li_4$ in IIA}

First we demonstrate that the sources have to vanish. If we compare
equation (\ref{bianchi1}) with equation (\ref{bianchi2}) and
(\ref{formeom6}), after substituting $\eta,g,e,\alpha,l$ and $h$
using (\ref{bianchi4})-(\ref{bianchi6}) and
(\ref{formeom1})-(\ref{formeom3}), we can deduce the following
\begin{equation}
\begin{split}
\beta \mu =0\,,\qquad q \mu = 0\,.
\end{split}
\end{equation}
Hence we can either take $\beta=0$ and $q=0$, which from
(\ref{Einstein2}) gives $z=1$, that is (A)dS solutions, or $\mu=0$.

This means that the charge has to be zero in order to find Lifshitz
solutions. This excludes the cosmological Lifshitz solutions in
ordinary massive IIA, since there is no ingredient to break the null
energy condition. Hence we discard the possibility of finding
solutions for $(\epsilon,\ve)=(-1,1)$.

\subsubsection*{Nogo for static $Li_4$ in IIA$^{*}$}

%The same type of reasoning is available when we might want to relax
%the relation (\ref{dW2}), such that $\d W_2$ and $\Omega_R$ are
%non-proportional. At a first glance this might produce new
%relations, but in fact will only create special cases for a few
%Bianchi identities.

Consider the following relation
\begin{align}
&6 a^2+2 f^2+6 g^2+6 h^2+2 m^2+8 q^2+6 s^2 +c^2 |W_2|^2+e^2 |W_2|^2+l^2|W_2|^2\nonumber\\
&\qquad +p^2 |W_2|^2+16 z+8 z^2+2 \alpha^2+2 \eta^2=0.
\end{align}
which is a combination of (\ref{dilaton}) and (\ref{Einstein1}),
with signs $(\epsilon,\ve)=(1,-1)$. This implies that $z\in (-2,0)$,
to have real solutions. Equation (\ref{Einstein3}) implies $z\in
(0,1)$ and hence provides a contradiction.

\subsubsection*{Nogo for static $Li_4$ in IIA}

In the case of $(\epsilon,\ve)=(1,1)$ we can find the following
relations
\begin{align}
q \left(8 \beta ^2+32 k^2+24 |W_1|^2+|W_2|^2\right)&=16 q z\,,\\
8 k^2+|W_2|^2+24+16 z+8 z^2+2\beta ^2 &=30 |W_1|^2\,,
%4 q^2+4-4 z+\beta ^2&=0.
\end{align}
which for $q\neq 0$ cannot be solved simultaneously. Furthermore we
can prove after some algebra that $q=0$ only 
gives imaginary solutions in any set of
signs, $(1,1)$ or $(-1,-1)$, with $z=0,-4$. Hence we can deduce that
the only possibility for a solution is for $q\neq 0$ and
$(\epsilon,\ve)=(-1,-1)$.

\subsubsection*{Solutions for cosmological $Li_4$ in IIA$^{*}$}

Assuming $\beta=0$, which emerges as a natural assumption when
reducing the equations, one finds the following relation
\begin{equation}
12 k^2+27 |W_1|^2+4 \left(3+4 z+z^2\right)=0.
\end{equation}
This implies that $z\in [-3,-1]$ where $z=-3$ or $z=-1$ have
$W_1=k=0$. The same equation therefore also excludes the possibility
for dS solutions in massive IIA$^{*}$.

Using this we are able to find infinite sets of solutions. One
particular simple solutions is in the set-up where the following
parameters are zero
\begin{equation}
W_1,\mu,\beta,k,c,p,\eta,g,e,\alpha,h
\end{equation}
Notice that only two Lorentz-breaking parameters are present.
Non-zero parameters are
\begin{equation}
|W_2|^2=48,\ z=-3,\ l=-1,\ q=2\,,
\end{equation}
and the rest are determined by the following equations
\begin{align}
0&=f s-a (m+2 s)\,,\\
0&=9 a^2+5 m^2+3 s^2-8f^2\,,\\
0&=2 f^2+3 s^2-2-m^2\,,
\end{align}
which have an infinite set of solutions. One instance is
\begin{equation}
f= -0.998921,\ s= 0.68206,\ a= -0.267852,\ m= 1.17954\,.
\end{equation}

\section{Solutions with running scalars in IIB$^{\star}$ and Euclidean IIB}
Let us drop the assumption that the scalar fields have to be
constant. For simplicity we consider the case where the vectors (and
tensors) are also turned off, as was done in \cite{Nakayama:2009gi}.
Instead of starting with a compactification Ansatz we first
investigate whether the following  Lagrangian in 4 dimensions
\begin{equation}
S=\int\d
x^4\sqrt{|g|}\Bigl(R-\tfrac{1}{2}G_{ij}\partial\phi^i\partial\phi^j
-\Lambda\Bigr)\,,
\end{equation}
can support Lifshitz geometries. The equations of motion read
\begin{align}
&R_{\mu\nu}=\tfrac{1}{2}G_{ij}\partial_{\mu}\phi^i\partial_{\nu}\phi^j
+ \tfrac{1}{2}\Lambda g_{\mu\nu}\,,\\
&
\frac{1}{\sqrt{|g|}}\partial_{\mu}(\sqrt{|g|}g^{\mu\nu}\partial_\nu\phi^i)+\Gamma^i_{jk}\partial^{\mu}\phi^j\partial_{\mu}\phi^k=0\,.
\end{align}
We make the assumption that the scalars only depend on the
holographic coordinate, $\phi^i=\phi^i(r)$. If we insist on having
the anisotropy parameters $a, b$ and $c$ not coinciding (since that
would correspond to (A)dS solutions), we infer from the Einstein
equations in the $x^i$ directions that
\begin{equation}
a+b+c=0=\Lambda\,,\qquad (z=-2)\,.
\end{equation}
In terms of the coordinate $\rho=\ln r$, the scalar field equation
of motion becomes the equation of motion for a geodesic on the
target space with metric $G_{ij}$ and with $\rho$ as affine
parameter
\begin{equation}
\ddot{\phi}^k +\Gamma^k_{ij}\dot{\phi}^i\dot{\phi}^j=0\,,
\end{equation}
where a dot represents a derivative with respect to $\rho$. Denoting
the constant affine velocity as $v^2$
\begin{equation}
G_{ij}\dot{\phi}^i\dot{\phi}^j=v^2\,,
\end{equation}
we can rewrite the $(rr)$ component of the reversed Einstein
equation in the following way
\begin{equation}
-(a^2+b^2+c^2)= \tfrac{1}{2}\,v^2\,.
\end{equation}
This shows that we need target spaces of indefinite signature such
that the geodesic velocity can be negative. In the case of Euclidean
field theories this can occur naturally. For instance, when IIB
supergravity is Wick-rotated it is known \cite{Gibbons:1995vg} that
the RR zero-form $C_0$ flips sign in the kinetic term, such that the
axion-dilaton part of the action in Einstein frame reads
\begin{equation}
S=\int\sqrt{g}\bigl(-\tfrac{1}{2}(\partial\phi)^2+\tfrac{1}{2}\e^{b\phi}(\partial
C_O)^2\bigr)\,,\label{sigma}
\end{equation}
where the number $b^2=4$ and represents the curvature of the
$\SL(2,\Real)/\SO(1,1)$ - sigma model. Upon dimensional reduction
this number $b$ grows and $\phi$ becomes a linear combination of the
the string coupling and the radii of the internal dimensions. In any
case, this demonstrates that the Euclidean Lifshitz solutions exist
in Euclideanised IIB supergravity or its dimensionally reduced
children. This solution closely resembles the so-named
\emph{non-extremal D(-1) solution or instanton}
\cite{Gutperle:2002km, Bergshoeff:2004fq, Bergshoeff:2005zf}. The
extremal D$(-1)$ solution corresponds to the case with lightlike
geodesics and the corresponding flat space metric
\cite{Gibbons:1995vg}.

Another example of indefinite kinetic terms is in II$^{\star}$
theories as we already discussed. In the case of IIB$^{\star}$
theory, we have the same sigma model (\ref{sigma}) without having to
Euclideanise the theory. This implies that we have the stationary
and the cosmological Lifshitz solutions in (dimensionally reduced)
IIB$^\star$ theory.

Finally, we like to emphasize that these solutions have vanishing
background cosmological constant. In case we add higher-derivatives
on the axion-dilaton the background cc does not vanish
\cite{Nakayama:2009gi} from the Einstein equations. In Euclidean IIB
theory this background cc in $D=5$ can be generated in the usual way
from the Freund-Rubin compactification (aka the D3 brane
near-horizon), also in the Euclidean case. Since the axion and
dilaton do not couple to the cc in the latter case we are in the
situation described by Nakayama \cite{Nakayama:2009gi}, however in
the Euclidean version of the theory. In the IIB$^{\star}$ case we
can generate the \emph{positive} background cc from the near-horizon
of the so-called E4 brane, or equivalently a Freund-Rubin
compactification on a (compactified) 5-dimensional hyperboloid with
F5 RR flux (where the F5 form has the opposite sign of the kinetic
term). This implies that $\alpha'$ corrections to our solutions
naturally allow a background cc.

\section{Discussion}
Let us summarize the results of this paper:
\begin{itemize}
\item We have shown that for a well-motivated and extended Ansatz one cannot
find static Lifshitz solutions in IIA on $\SU(3)$-structure
manifolds (and orientifolds). \item However, the same Ansatz does
allow explicit cosmological Lifshitz solutions in IIA$^*$
supergravity. For simplicity we presented a solution with $z=-3$,
but there are other possible solutions  with $z$-values between
$[-3,-1)$.
\item When we allow running scalar fields we can construct static
and cosmological solutions with $z=-2$ in IIB$^*$ supergravity and a
Euclidean Lifshitz solution in Euclidean IIB supergravity, also with
$z=-2$, that can be interpreted as a non-extremal deformation of the
D-instanton.
\end{itemize}

The nogo conditions we obtained for the static Lifshitz backgrounds
needs some interpretation in order to be useful. We can think of two
scenarios. Either we cannot find a solution of the Ansatz with
constant scalars simply because the Ansatz is restricted in many
ways. It assumes $\SU(3)$-structures, and more importantly, it
assumes fluxes along specific directions. Especially the
Lorentz-violating terms have to be chosen with care since they
correspond to the 4-dimensional massive tensors and vectors, see e.g. \cite{Cassani:2009ck}. It
could be that we have turned on those massive vectors and tensors
that do not lead to a solution\footnote{We thank Yu Nakayama for
some explanations on that point.}. An alternative explanation is
that we do not find solutions to the Ansatz because of the
assumption of constant scalar fields. We can schematically write the
4D effective theory in the following way:
\begin{equation}
S=\int\sqrt{g}\bigl\{R - \tfrac{1}{2}(\partial\phi)^2 - f_1(\phi)F^2
-f_2(\phi)m^2A^2 -V(\phi) \bigr\}\,,
\end{equation}
where we pretended, for simplicity, that only one vector and one
scalar is turned on. $f_1$ and $f_2$ are some functions of the
scalar and $V$ is the scalar potential. Since we consider AdS
compactifications the scalar potential has a stationary point at a
negative value
\begin{equation}
V(\phi_*)<0\quad \& \quad \partial V_{|\phi_*}=0\,.
\end{equation}
If the vector is non-zero, the scalar field equation of motion
effectively feels a new scalar potential $\tilde{V}$, where
\begin{equation}
\tilde{V}(\phi)=V(\phi) +\alpha f_1(\phi) +\beta f_2(\phi)
\end{equation}
with $\alpha, \beta$ some real numbers. There is no guarantee that
$\tilde{V}$ has also a stationary point, which would imply that the
scalar field has to run. The easiest way to check this possibility
is by investigating the effective field theories in four dimensions
directly.

For the cosmological Lifshitz solutions in IIA$^*$ theory we found
that $V$ has no stationary point, since there are no dS solutions in
our model, but $\tilde{V}$ has since we did find the Lifshitz
solutions. This implies that one does not necessarily have to
consider AdS/dS compactifications in order to find
static/cosmological Lifshitz solutions. In such cases there will not
be a partner AdS/dS solution to the static/cosmological Lifshitz
solution.

For both aforementioned reasons we anticipate on investigating the
existence of static Lifshitz solutions in 4d effective field
theories. One could even relax the requirement of knowing the 10d
origin of the effective field theories and consider that question as
a second step. A sensible set of theories to investigate would be
gauged $\mathcal{N}=2$ supergravities coupled to massive tensor
multiplets. Such theories have the necessary ingredients for
Lifshitz solutions and are still constrained enough to make the
analysis tractable. These theories also are expected to originate
from generalised Calabi-Yau flux compactifications of 10d
supergravity.

Concerning the Euclidean Lifshitz solution with $z=-2$ we have found
in IIB (as well as the static and cosmological IIB$^*$ solutions) we
have not yet touched upon issues regarding the regularity of the
solution. Generically the dilaton profile has a singularity in its
derivative as well as the axion field \cite{Bergshoeff:2004fq,
Bergshoeff:2005zf}. However, this cancels in the Einstein equation
such that this singular point does not backreact on the geometry.
Nonetheless, it might be an issue of concern when taking the
solutions serious. We will not go into this discussion here but
mention a possible way out of the problem. Upon going to Euclidean
signature one might allow axionic fields different from $C_0$ to
Wick-rotate as well. The effect of the multiple axion-dilaton pairs
then removes the singularity \cite{Arkani-Hamed:2007js,
Bergshoeff:2008be}. Another option is to look at more involved
compactifications such that one ends up with $\mathcal{N}=2$
Euclidean theories, where the singularity is also absent
\cite{Bergshoeff:2004pg}.

\section*{Acknowledgements}
We thank Yu Nakayama for very useful correspondence and Paul Koerber
for useful discussions on generalised complex geometry. T.V.R. is
supported by the G\"{o}ran Gustafsson Foundation. U.D. is supported
by the Swedish Research Council (VR) and the G\"{o}ran Gustafsson
Foundation.

\bibliography{Lifshitz}
\bibliographystyle{utphysmodb}
\end{document}